\documentclass[aps,prb,nofootinbib,twocolumn,amsmath,amssymb]{revtex4}
\usepackage{graphicx}
\usepackage{bm}
\usepackage[breaklinks=true,colorlinks,citecolor=blue,linkcolor=blue,urlcolor=blue]{hyperref}

\pdfpageattr{/Group << /S /Transparency /I true /CS /DeviceRGB>>}

\begin{document}
% TITLE
\title{Emergence of one-dimensional wires of free carriers in transition-metal-dichalcogenide nanostructures}
%AUTHORS
\author{Marco Gibertini}
\email{marco.gibertini@epfl.ch}
\author{Nicola Marzari}
\email{nicola.marzari@epfl.ch}
\affiliation{Theory and Simulation of Materials (THEOS) and National Centre for Computational Design and Discovery of Novel Materials (MARVEL), \'Ecole Polytechnique F\'ed\'erale
de Lausanne, CH-1015 Lausanne, Switzerland}
% ABSTRACT
\begin{abstract}
We highlight the emergence of metallic states in two-dimensional transition-metal-dichalcogenide nanostructures --nanoribbons, islands, and inversion domain boundaries-- as a widespread and universal phenomenon  driven by the polar discontinuities occurring at their edges or boundaries. 
We show that such metallic states form one-dimensional wires of electrons or holes, with a free charge density that increases with the system size, up to complete screening of the polarization charge, and can also be controlled by the specific edge or boundary configurations, e.g. through chemisorption of hydrogen or sulfur atoms at the edges. 
For triangular islands, local polar discontinuities occur even in the absence of a total dipole moment for the island and lead to an accumulation of free carriers close to the edges, providing a consistent explanation of previous experimental observations. 
To further stress the universal character of these mechanisms, we show that polar discontinuities give rise to metallic states also at inversion domain boundaries.
These findings underscore the potential of engineering transition-metal-dichalcogenide nanostructures for manifold
applications in nano- and opto-electronics, spintronics, catalysis, and solar-energy harvesting.
\end{abstract}

\maketitle

Transition metal dichalcogenides (TMDs) often crystallize in a layered structure so that  two-dimensional (2D) mono- or few-layer sheets can be easily extracted from bulk materials, e.g. through exfoliation, similarly to what happens for graphene from graphite.
The electronic and structural properties of such 2D TMDs --with chemical formula MX$_2$-- depend crucially on the transition metal atom M and on the chalcogen X ($=$S, Se, or Te) involved\cite{Chh2013}. Recently, particular emphasis has been devoted to group-VI TMDs where M$=$Mo or W\cite{Chh2013,Wang2012}. These 2D materials are semiconductors (with the only exception of the semimetal WTe$_2$) and offer the potential for extremely interesting technological applications\cite{Chh2013,Wang2012,Jar2014,Lv2015} in electronics, optoelectronics, (pseudo-)spintronics, photonics, and plasmonics.   

When designed into nanostructures like nanoribbons or triangular islands, group-VI TMDs become metallic at the edges, while remaining semiconducting in the bulk, owing to the appearance of edge states that cross the bulk energy gap. Such metallic edge states have been reported both theoretically\cite{Bol2003,Li2008,Voj2009,Bot2009,Ata2011,Pan2012,Pan2012b} and experimentally\cite{Hel2010,Bol2001} along zigzag edges. Their presence is associated with an enhanced photoluminescence response\cite{Gut2013} and with an extraordinary catalytic behavior\cite{Lau2003}, both in hydrodesulfurization\cite{Lau2004,Chi2006} and hydrogen evolution reactions\cite{Hin2005,Jar2007}. In addition, it has been theoretically predicted that these metallic edge states are magnetic\cite{Li2008,Voj2009,Bot2009,Ata2011,Pan2012,Pan2012b} with promising applications in spintronics and, more recently, that they give rise to unusual one-dimensional (1D) plasmonic excitations\cite{And2014}. Interestingly, there are both theoretical\cite{Zou2013,Zhou2013} and experimental\cite{Liu2014} evidences that similar metallic states appear also at inversion domain boundaries extending along a zigzag direction.

The existence of such metallic states at zigzag edges in group-VI TMD nanoribbons has been predicted to be a consequence of the non-zero bulk polarization of these materials\cite{Gul2013,Gib2014}. 
Indeed, a polar discontinuity appears at the edges of nanoribbons with the appearance of localized polarization charges. These charges  give rise to an electric field that induces a charge reconstruction consisting in the accumulation of free carriers at the edge in order to screen the polarization charges and prevent a \emph{polar catastrophe}\cite{Nak2006}.   This mechanism underlies similar insulator-to-metal transitions in SrTiO$_3$/LaAlO$_3$ interfaces\cite{Pent2010,Bris2014} and their 2D counterparts\cite{Gib2014,Bris2013}, and has important implications in graphene-boron nitrid heterostructures\cite{Pru2010}.

In this paper we address metallicity in group-VI TMD nanostructures in order to attain full control over their possible technological applications. 
With the help of extensive first-principles density-functional-theory (DFT) calculations, we investigate how the charge density of free carriers varies with the system size and we explore the effects of specific edge/boundary configurations. In particular, we show that in some cases additional bound charges appear and affect the asymptotic value of the free charge density; as a paradigmatic example we investigate here the case of  2D TMD nanoribbons terminated with hydrogen or sulfur atoms, where metallic edge states are still present\cite{Bol2003,Pan2012}.  We show that the existence of a polar discontinuity is at the heart of the metallic states appearing also in triangular islands and inversion domain boundaries.
Last, such extensive investigation of the peculiar nature of these edge/boundary states paves the way for new exciting applications in solar-energy devices.

\begin{figure}
\includegraphics{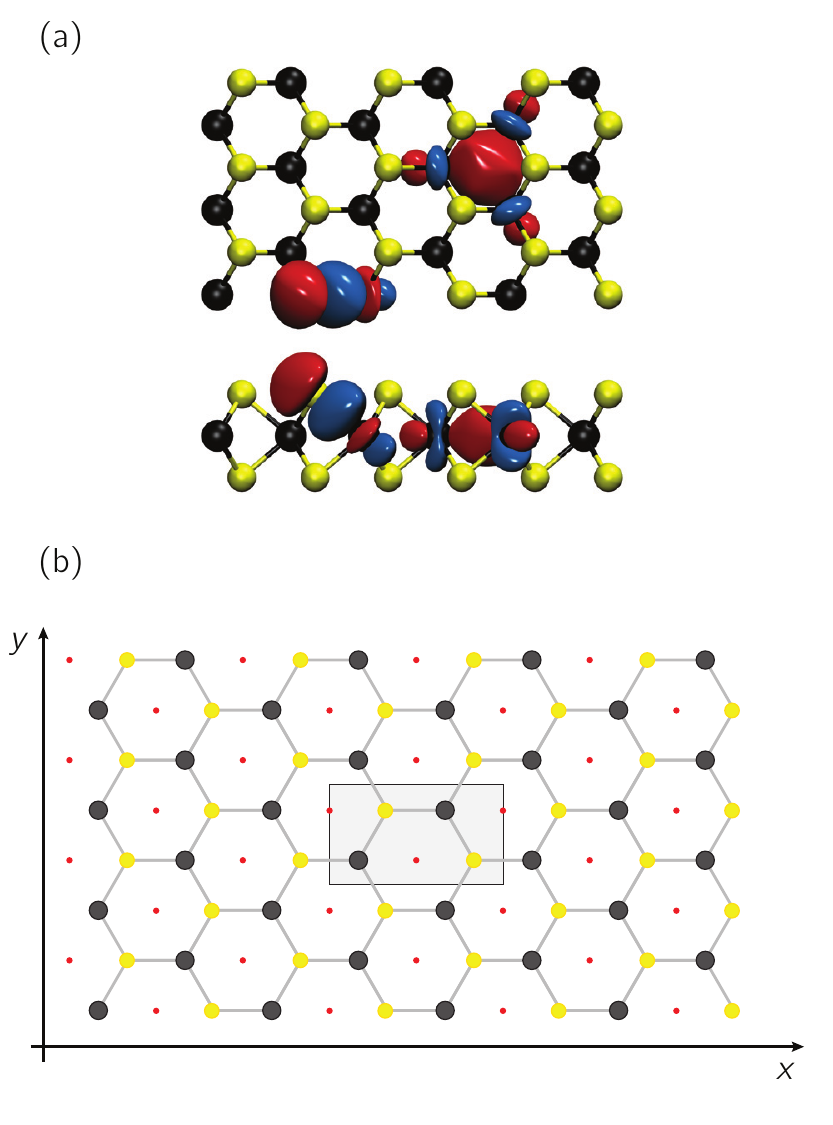}
\caption{Panel (a): Top and lateral views of the crystal structure of group-VI 2D TMDs  (here MoS$_2$) with isosurface plots of the two kinds of valence Wannier functions: (i) a $p$-like Wannier function centered on a chalcogen atom (yellow),  hybridized with transition-metal $d$-orbitals; and (ii) a Wannier function centered in the middle of the hexagonal cell, with contributions from the $d$-orbitals of the transition metal atoms (black) surrounding it. Panel (b): Schematic representation of the crystal honeycomb structure where in addition to chalcogen (yellow) and transition metal (black) atoms also the centers of the Wannier function (ii) above are reported (small red circles). A non-primitive rectangular unit cell is emphasized with gray shading. When repeated periodically along the $y$-axis, this structure give rise to a $N_{\rm w}=4$ zigzag nanoribbon, while when repeated along $x$ it leads to a $N_{\rm w}=4$ armchair nanoribbon. 
\label{fig:wannier}}
\end{figure}

Monolayer group-VI TMDs have an hexagonal lattice with three atoms (MX$_2$) in the unit cell. The atoms are arranged in three parallel monoatomic planes X-M-X, with transition-metal and chalcogen atoms forming two distinct 2D triangular sublattices (see Fig.~\ref{fig:wannier}). This gives each M atom a trigonal prismatic coordination and the overall point group is $D_{3h}$. As already mentioned above, all group-VI TMDs are semiconductors (the only exception being the semimetal WTe$_2$, for which the most stable structure is not the one described above), with a direct band gap at the two inequivalent corners of the Brillouin zone. 
The bulk polarization\cite{King1992,Resta1994} of 2D TMDs can be conveniently computed by mapping the ground state of the system into a set of maximally-localized Wannier functions\cite{Mar2012}.
Then, the electronic contribution to the polarization is
a sum over point-like charges located at the Wannier centers $\langle\bm{r}\rangle_{j}$
and the total polarization reads
\begin{equation}
\bm{P}=\frac{e}{\Sigma}\left(\sum_{\alpha=1}^{N}Z_{\alpha}\bm{\tau}_{\alpha}-2\sum_{j=1}^{N_{{\rm el}}/2}\langle\bm{r}\rangle_{j}\right)+\frac{2e}{\Sigma}\bm{R}\,.\label{eq:pol_wannier}
\end{equation}
Here $Z_{\alpha}$ and $\bm{\tau}_{\alpha}$ are the charges and positions
of the $N$ ions in the unit cell, $N_{{\rm el}}$ is the number
of electrons,  and $\Sigma$ is the area of a unit cell. In Eq.~(\ref{eq:pol_wannier}) $\bm{R}$ is a generic Bravais lattice vector, since in the modern theory of polarization\cite{King1992,Resta1994} the formal polarization is not a single vector (like in classical electrostatics) but rather a lattice of vectors. If we consider as valence electrons only the outermost $p$-electrons of the chalcogens and the  $s$- and $d$-electrons of the transition metal, we then have that $Z_{\rm X}=4$, $Z_{\rm M}=6$, and $N_{\rm el}=14$.
In the following, maximally localized Wannier functions corresponding to the valence bands of 2D TMDs have been computed using the Wannier90 code\cite{Wannier90} starting from the DFT ground state  obtained with Quantum-ESPRESSO\cite{Gia2009} (see Methods for more details). 
Owing to their isovalence and isostructural properties, the valence Wannier functions have the same character for all group-VI 2D TMDs (in the case of WTe$_2$ we consider a metastable semiconducting structure identical to that of the other group-VI TMDs). In particular, we find three $p$-like Wannier functions centered on each chalcogen,  hybridized with transition-metal $d$-orbitals, together with a Wannier function centered in the middle of the hexagonal cell, with contributions from the $d$-orbitals of the atoms surrounding the cell\cite{Gib2014} (see Fig.~\ref{fig:wannier}(a)). With this set of valence Wannier functions and their centers it is  straightforward to compute the bulk formal polarization according to Eq.~(\ref{eq:pol_wannier}). For convenience we consider a non-primitive rectangular unit cell, as shown in Fig.~\ref{fig:wannier}, with the long side along an armchair direction ($x$-axis) and the short side along a zigzag direction ($y$-axis). 
The bulk formal polarization is then given by
\begin{equation}\label{eq:pol_tmd}
\bm{P}=-\frac{2e}{3a}\hat{\bm{x}} \,,
\end{equation}
where $a$ is the lattice constant and we set $\bm{R}=0$ in Eq.~(\ref{eq:pol_wannier}), thus selecting a specific representative element of the polarization lattice associated with the choice of the unit cell in Fig.~\ref{fig:wannier}~\cite{Vand1993,Ste2011}. We stress that $\bm{P}$ is quantized and points  along an  armchair direction, in agreement with the constraints imposed by the three-fold rotation symmetry\cite{Fang2012,Jad2013}.

In a nanoribbon the periodicity of the bulk crystal is preserved along a given direction, which defines the direction of the nanoribbon, while it is broken in the orthogonal direction where the edges create a discontinuity. Since vacuum can be considered as an insulator with zero polarization, we have that across the edge of a group-VI TMD nanoribbon a polar discontinuity arises. Polarization charges will thus appear with a linear charge density given by 
\begin{equation}\label{eq:polcharge}
\lambda_{\rm P} = \bm{P}\cdot\hat{\bm{n}}\,,
\end{equation}
where $\hat{\bm{n}}$ is a unit vector orthogonal to the edge. 
By combining Eq.~(\ref{eq:pol_tmd}) and (\ref{eq:polcharge}), we see that when a nanoribbon extends along a zigzag direction, the polarization is orthogonal to the edge, so that the polarization charge is maximal and it is given by $\lambda_{\rm P}=\pm 2e/(3a)$, where the plus sign refers to the Mo-edge while the minus sign to the S-edge. For armchair nanoribbons instead the polarization is parallel to the edge and the polarization charge vanishes ($\lambda_{\rm P}=0$). 
This difference in the relative orientation between the edges and the bulk polarization is at the heart of the different electronic properties reported in the literature for  zigzag and armchair TMD nanoribbons\cite{Bol2003,Li2008,Bot2009,Ata2011,Pan2012,Pan2012b}. Indeed, as we will show below, it is the electric field associated with the polarization charges that triggers an insulator-to-metal transition and drives the appearance of metallic edge states in zigzag (but not in armchair) nanoribbons. 

\begin{figure}
\includegraphics{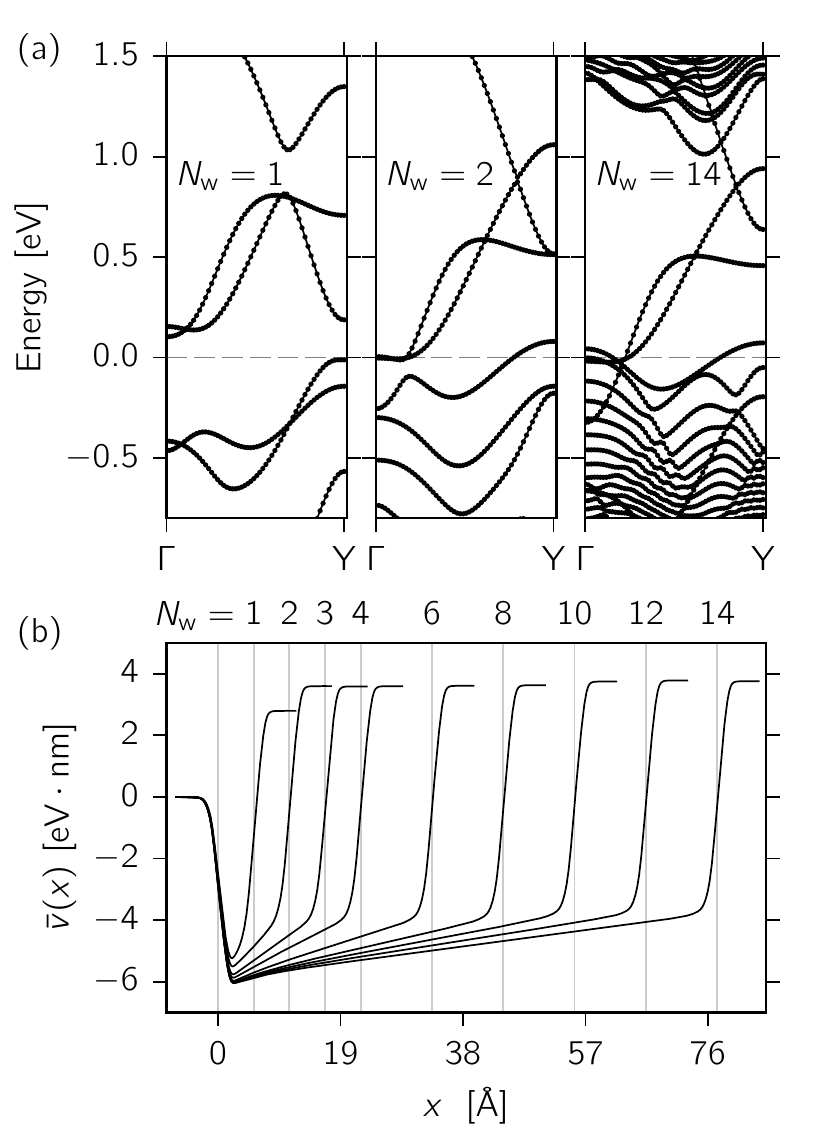}
\caption{Panel (a): Band structures of zigzag MoS$_2$ nanoribbons with different widths: $N_{\rm w}= 1,2$, and 14. The dashed line highlights the position of the Fermi energy. Panel (b): Macroscopic average of the electrostatic potential energy defined in Eq.~(\ref{eq:avpot}) as a function of the coordinate across the ribbon. Results are reported for  nanoribbon widths $N_{\rm w}=1,2,3,4,6,8,10,12,14$. 
\label{fig:znr}}
\end{figure}

\begin{figure}
\includegraphics{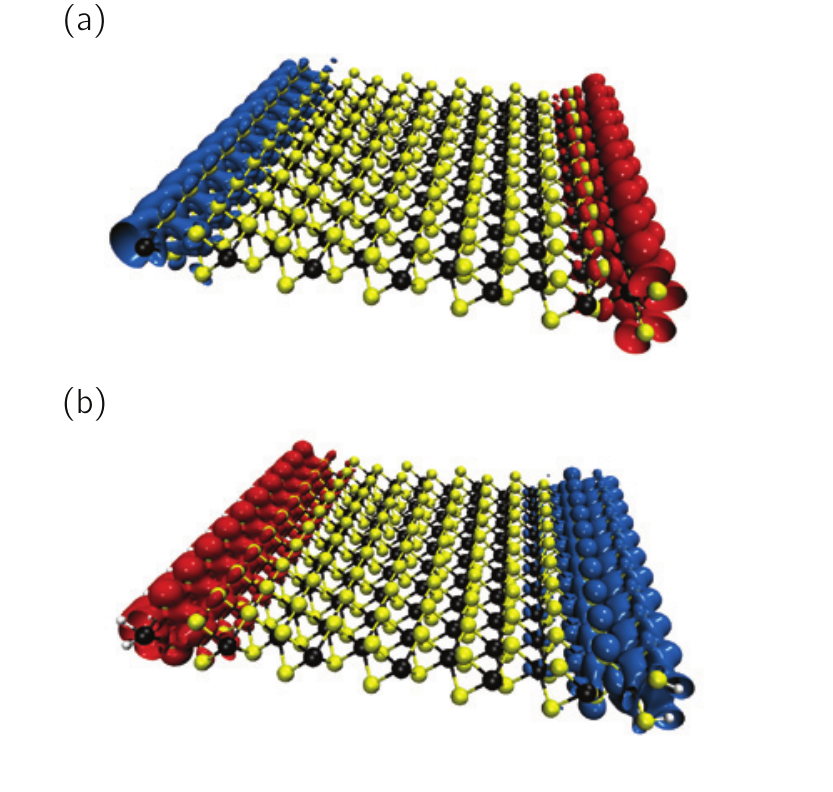}
\caption{Real-space distribution of the charge density of free electrons (blue) and holes (red) for a $N_{\rm w}=6$ zigzag MoS$_{2}$ nanoribbon with (a) bare and (b) hydrogen-terminated  edges. \label{fig:chargedens}}
\end{figure}

Let us first consider zigzag nanoribbons, obtained by the repetition of the rectangular unit cell shown in Fig.~\ref{fig:wannier}(b) along the $x$-direction $N_{w}$ times, while being periodic along the $y$-direction.  In Fig.~\ref{fig:znr}(a) we report the band structure of MoS$_2$ nanoribbons with different widths, corresponding to $N_{w}  =$ 1, 2, and 14, respectively. Qualitatively similar results can be obtained for all other 2D TMDs. 
For very narrow nanoribbons ($N_{w}$=1) the system is semiconducting, although the energy band gap is significantly reduced with respect to the bulk owing to the presence of mid-gap states that are localized at the edges. For larger ribbon widths, the gap closes and the system becomes metallic with a charge transfer from the top valence band to the bottom conduction bands. To understand this insulator-to-metal transition, we also show in Fig.~\ref{fig:znr}(b)  the macroscopic average\cite{Bal1988} $\bar v (x)$ of the electrostatic potential energy $v(x,y,z)$, defined as
\begin{equation}\label{eq:avpot}
\bar v (x) = \frac{1}{\Sigma} \int_{x-\sqrt{3}a/4}^{x+\sqrt{3}a/4}\int_{0}^{a}\int_{-\infty}^{\infty} v(x-x^\prime,y,z) dx dy dz.
\end{equation}
The presence of polarization charges at the edges is confirmed by the change in slope of $\bar v(x)$ across each edge. Indeed, although this slope is not related to the actual electric field acting on the electrons owing to the integral over an infinite range in the definition (\ref{eq:avpot}), Gauss theorem still applies and the change in slope of $\bar v(x)$ gives an important information on the presence of a finite amount of total charge at the nanoribbon boundaries and thus on the existence of polarization charges. 
These polarization charges induce an electric field that in turn gives rise to a potential energy difference $\Delta v$ between the two edges of the nanoribbon. This means that valence states on one edge are pushed upwards in energy, while conduction states at the opposite edge are pushed downwards, with an overall reduction of the global energy gap of the system. As the width of the nanoribbon increases, $\Delta v$ grows logarithmically\cite{Gul2013}, with a progressive decrease of the gap. Above a critical width, the gap vanishes and the system becomes metallic, thus explaining the evolution of the band structure provided in Fig.~\ref{fig:znr}(a). This insulator-to-metal transition as a function of the nanoribbon width is associated with a \emph{charge reconstruction}: electrons from the top valence bands are transferred to the bottom conduction bands with the creation of electron and hole pockets of free carriers. In Fig.~\ref{fig:chargedens}(a) we show the spatial profile of the charge density of free carriers for a $N_{w} = 6$ MoS$_2$ nanoribbon, as obtained by integrating the local density of states associated with the pockets of free electrons and holes.  We note that these free carriers form 1D metallic wires extending  along the nanoribbon edges and present opposite character (electron or hole) at opposite edges: electrons on the Mo-edge and holes on the S-edge. 
In the following we define as $\lambda_{\rm F}$ the total free charge (per unit length) present in such 1D wires, specifying wherever necessary if we refer to the chalcogen or transition-metal  edge.

\begin{figure}
\includegraphics{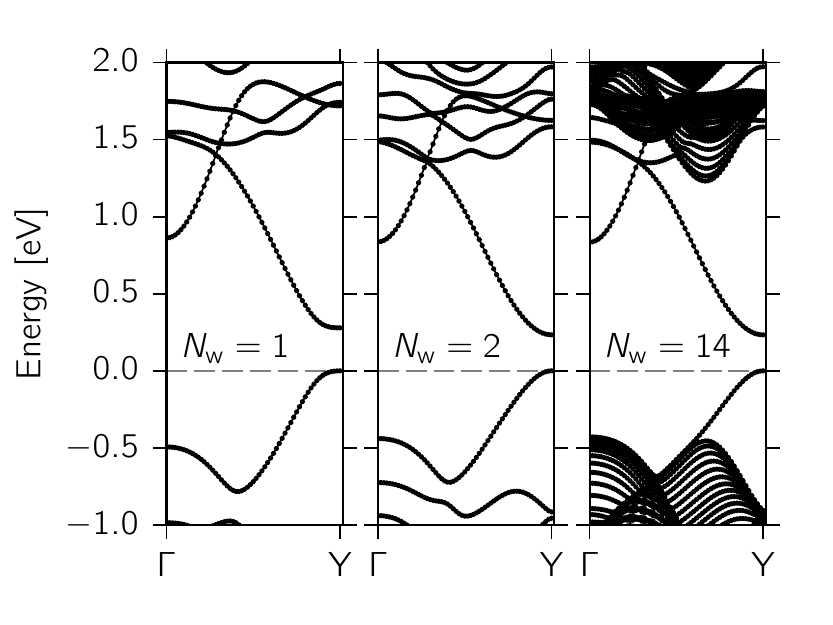}
\caption{Band structures of zigzag MoS$_2$ nanoribbons with different widths ($N_{\rm w}= 1,2$, and 14). Differently from Fig.~\ref{fig:znr}(a), in this case the energy bands are obtained from a tight-binding model that does not include electrostatic effects. 
\label{fig:tb}}
\end{figure}

Further evidence of the electrostatic origin of metallic edge states is given in Fig.~\ref{fig:tb}, where we show the  band structures of MoS$_2$ zigzag nanoribbons with different widths obtained using a tight-binding model. 
The tight-binding parameters have been obtained for a bulk monolayer MoS$_2$ by mapping both the valence and the conduction bands into a set of atom-centered Wannier functions corresponding to Mo $d$-orbitals and S $p$-orbitals\cite{Gib2014b}. Although edge states are still present owing to the undercoordination of edge atoms, the system remains always semiconducting with a gap that does not change with the nanoribbon width. As a consequence, in this case there is no accumulation of free carriers at the edges. Since electrostatic effects are not included within this tight-binding description, this means that, although polarization charges and their corresponding electric field are not necessary to account for the existence of edge states in TMD nanoribbons, they are needed to drive metallicity. 
In addition, comparing in more detail the band structures reported in Figs.~\ref{fig:znr} and \ref{fig:tb} we observe that the electric field leads also to a relative shift between the energy bands. This effect is particularly relevant on the Mo-edge where the electrons that, according to the bulk calculations, should occupy the Wannier function centered outside the ribbon (see also Fig.~\ref{fig:wannier}(b)) feel the attraction of the positive polarization charges and are thus transferred on the Mo-atom. This brings a band, which according to the tight-binding model would be unoccupied, below the Fermi energy and correspondingly a tight-binding valence band is pushed above the Fermi energy. The charge reorganization associated with the change in the band occupation can be effectively described in terms of an additional $+2e$ bound charge at the original position of the Wannier center and a $-2e$ charge on the external Mo atom, so that the overall charge at the edge remains the same. 

\begin{figure}
\includegraphics{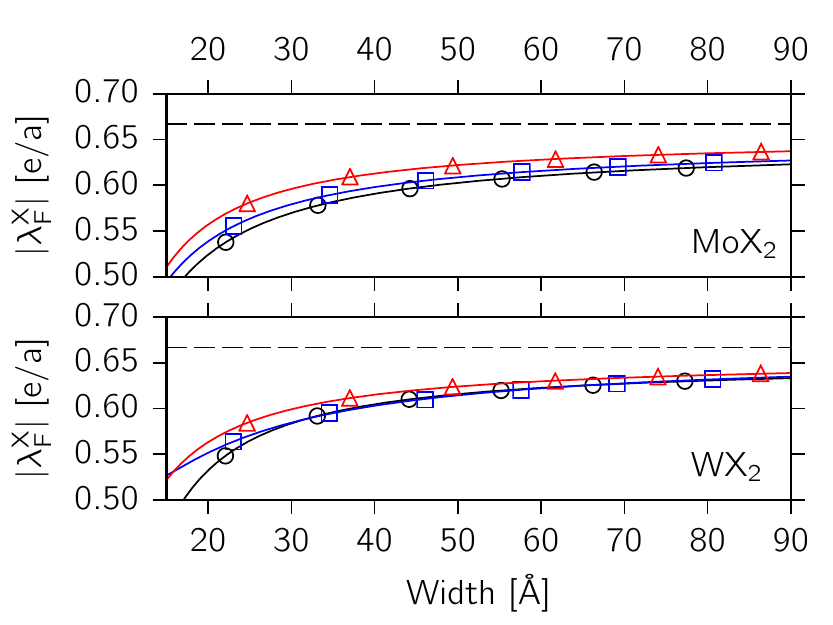}
\caption{Charge density of free carriers $|\lambda^{\rm X}_{\rm F}|$ at the chalcogen-edge as a function of the nanoribbon width for: (a) Mo-based and (b) W-based group-VI transition metal dichalcogenides. Different symbols identify different chalcogen atoms: S (black circles), Se (blue squares), and Te (red triangles).  In all cases the free charge density asymptotically reaches the value $2e/(3a)$ (dashed line), corresponding to perfect screening of the polarization charge $\lambda_{\rm P}$. Lines are fits according to Ref.~\onlinecite{Gul2013}.
\label{fig:free-ribbon}}
\end{figure}

It is important to mention that the free charges associated with the edge states partially screen the polarization charges, so that the total amount of charge at each edge decreases when the system becomes metallic. Indeed, in Fig.~\ref{fig:znr}(b)  we observe that the slope of $\bar v(x)$ drops when the nanoribbon from semiconducting ($N_{w}=1$) becomes metallic ($N_{w}=2$). In addition, the slope further decreases as the nanoribbon width increases, suggesting that the charge density of free carriers $\lambda_{\rm F}$ 
grows with $N_{w}$. This is due to the fact that, as the nanoribbon width increases, the electric field drives more and more free charges to the edges, until (asymptotically) these perfectly screen the polarization charges and the electric field vanishes. 
To further confirm this picture, in Fig.~\ref{fig:free-ribbon} we show $|\lambda_{\rm F}|$ at the chalcogen edge as a function of the nanoribbon width for all group-VI TMD. In agreement with the above analysis on the slope of $\bar v(x)$, it can be seen that the free charge increases with the nanoribbon width and asymptotically it equals the polarization charge, $|\lambda_{\rm F}| = |\lambda_{\rm P}|=2e/(3a)$. Qualitatively similar behavior can be observed for all group-VI TMDs, with only minor quantitative differences in the rate at which the asymptotic limit is reached. 
Differences emerge instead when considering the free charge on the transition metal edge. Indeed, while in all Mo-based TMD nanoribbons the free charge on the Mo-edge is identical (in magnitude) to the one on the chalcogen edge,  for W-based nanoribbons we have observed that the free charge on the W-edge exceeds by $2e/a$ the free charge on the chalcogen edge for all widths. This is due to the fact that also the band associated with the extra electrons on the external W-atom becomes metallic in WX$_2$ nanoribbons. 

\begin{figure}
\includegraphics{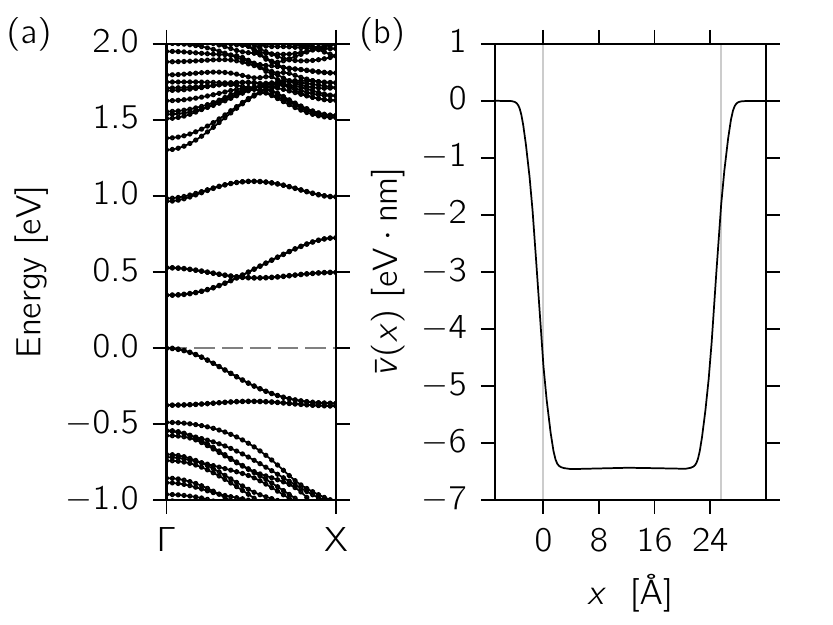}
\caption{(a) Energy band structure  and (b) profile of the macroscopic average of the electrostatic potential energy   for a $N_{\rm w}=8$ MoS$_2$ armchair nanoribbon.
\label{fig:arm}}
\end{figure}

In Fig.~\ref{fig:arm} we summarize what happens for armchair nanoribbons that result from the $N_{w}$-fold repetition of the rectangular unit cell shown in Fig.~\ref{fig:wannier} along the $y$-direction, while preserving the bulk periodicity along the $x$-direction. In agreement with previous literature\cite{Li2008,Bot2009,Ata2011,Pan2012,Pan2012b} we find that such nanoribbons remain semiconducting at all widths, although the energy gap is reduced with respect to the bulk owing to the presence of mid-gap states associated with the undercoordination of the edge atoms. Indeed, in this case an insulator-to-metal transition does not occur owing to the absence of polarization charges at the edges. This is most evident in Fig.~\ref{fig:arm} where we see how the macroscopic average of the electrostatic potential $\bar v(x)$ remains flat inside the nanoribbon, meaning that no net charge is present at the edge.

\begin{figure*}
\includegraphics{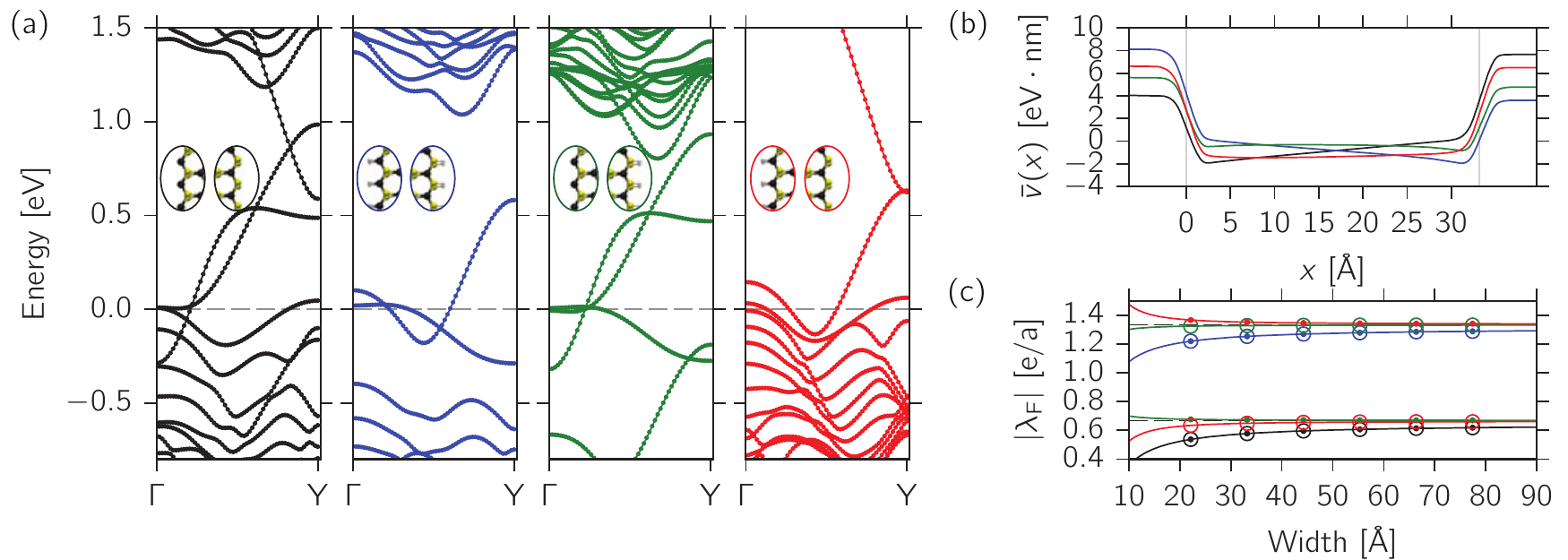}
\caption{Panel (a): Energy bands for MoS$_2$ nanoribbons ($N_{\rm w}=6$) with different edge terminations: bare edges (black), hydrogen-terminated edges (blue), hydrogen-terminated S-edge (green), and hydrogen-terminated Mo-edge (red).
Panel (b): Macroscopic average of the electrostatic potential energy defined in Eq.~(\ref{eq:avpot}) as a function of the coordinate across the ribbons introduced in panel (a). 
Panel (c): Charge density of free carriers as a function of the nanoribbon width for different edge terminations. Large empty circles represent the charge density on the S-edge, $|\lambda^{\rm S}_{\rm F}|$, while small filled circles refer to the Mo-edge, $|\lambda^{\rm Mo}_{\rm F}|$. Depending on the edge termination, two different asymptotic limits are reached, at $2e/(3a)$ or $4e/(3a)$ (dashed lines).
Lines are fits according to Ref.~\onlinecite{Gul2013}.
The color coding in panels (b) and (c) is the same as that in panel (a).
\label{fig:hydrogen}}
\end{figure*}

Up to this point we have assumed perfect zigzag or armchair ribbons where the edge atoms remain undercoordinated. In the following we would like to address termination effects, when additional atoms are chemically bonded at the zigzag edges to restore  bulk coordination. For definiteness we focus on MoS$_2$ and consider hydrogen or sulfur as saturating the edge dangling bonds, since these are species that are typically present during growth.  In the case of hydrogen, we assume to have two H atoms bonded to the external molybdenum on the Mo-edge in order to recover  bulk  six-fold coordination, while at the S-edge each sulfur atom binds to a single H atom. 
We consider separately each edge since these can be assumed decoupled for sufficiently large nanoribbon widths. At the Mo-edge, in order to form a bond with the hydrogen atoms two electrons are needed. This means that the total amount of bound charge $\lambda_{\rm b}^{\rm Mo}$ at the Mo-edge is not simply given by the polarization charge $\lambda_{\rm P}^{\rm Mo}$, but there must be an additional $-2e$ contribution per unit length: $\lambda_{\rm b}^{\rm Mo}=\lambda_{\rm P}^{\rm Mo} -2e/a = -4e/(3a)$. Similarly, on the S-edge the S-H bond is partially ionic, with sulfur removing an electron from the hydrogen. As a consequence, on this edge there is an additional $+2e$ bound charge per unit length due to the hydrogen ions, and the total bound charge on the S-edge is thus given by $\lambda_{\rm b}^{\rm S} = \lambda_{\rm P}^{\rm S} + 2e/a = 4e/(3a)$.   
Hence, we notice that with respect to pristine edges hydrogen termination leads to a change in the sign of the bound charge on both edges. For narrow nanoribbons, these bound charges are not compensated by free charges and create an electric field which points in the  opposite direction with respect to pristine edges. In Fig.~\ref{fig:hydrogen}(b) we show the macroscopic average of the electrostatic potential energy $\bar v(x)$ (blue line), confirming that the slope is inverted with respect to the case of pristine edges (black line), signaling a change in the sign of the bound charges in agreement with the discussion above. The ensuing electric field will drive an insulator-to-metal transition as it was the case for pristine edges, and for sufficiently large ribbons the system becomes metallic, as shown in Fig.~\ref{fig:hydrogen}(a) where we plot the band structure for a MoS$_2$ zigzag nanoribbon with hydrogen-terminated edges.  The only difference with respect to the pristine case is that now there are free holes on the Mo-edge and free electrons on the S-edge. This is reported in Fig.~\ref{fig:chargedens}(b) where we plot the charge density of free carriers for hydrogenated edges, and show that electrons and holes are exchanged with respect to Fig.~\ref{fig:chargedens}(a).  
Asymptotically, the free charge has to screen the bound charge in order to prevent a divergency in the electrostatic energy. In Fig.~\ref{fig:hydrogen}(c) we show the free charge density $|\lambda_{\rm F}|$ as a function of the nanoribbon width confirming that asymptotically there is perfect screening of the bound charge: $|\lambda_{\rm F}| \to |\lambda_{\rm b}^{\rm Mo}| =  |\lambda_{\rm b}^{\rm S}| = 4e/(3a)$. In summary, we have shown that the presence of hydrogen on both edges does not prevent the occurrence of an insulator-to-metal transition and of a finite electric field inside the nanoribbon, although its sign is reversed with respect to pristine edges and consequently also the character of free carriers along the 1D wires at the edges is exchanged. 

Before discussing the case of S-termination we consider the case in which hydrogen atoms are added only at the Mo- or at the S-edge. According to the analysis above, in these cases we have that the bound charges have the same sign on both edges but different values. For instance, when hydrogen is present only at the Mo-edge we have that the bound charge is always negative and given by $\lambda_{\rm b}^{\rm Mo} = -4e/(3a)$ on the Mo-edge and by $\lambda_{\rm b}^{\rm S}=-2e/(3a)$ on the S-edge. Analogously, when only the S-edge is terminated with hydrogen, the bound charge is always positive and is $\lambda_{\rm b}^{\rm Mo} = 2e/(3a)$ on the Mo-edge and $\lambda_{\rm b}^{\rm S}=4e/(3a)$ on the S-edge. In order to screen the electric fields associated with these bound charges, a charge reconstruction occurs and free carriers appear with the same character on both edges. The system is thus metallic, as shown in Fig.~\ref{fig:hydrogen}(a) where we report the band structures both in the case of hydrogen only on the Mo- or on the S-edge. In Fig.~\ref{fig:hydrogen}(c) we show the charge density of free carriers as a function of the nanoribbon width and we note that for sufficiently large widths the charge density of free carriers is constant and identical to the bound charge density and it is thus different on the two edges. This means in particular that at such nanoribbon widths the total net charge at each edge is zero and there is not an overall electric field. Indeed, in Fig.~\ref{fig:hydrogen}(b) the slope of $\bar v(x)$ stays flat across the ribbon when hydrogen is present only on the S-edge (green line) or on the Mo-edge (red line).  
Thus, in the hybrid situation in which only one edge is terminated with hydrogen we no longer have an overall electric field across the ribbon but we still have metallic edge states. 
Only for very small nanoribbon widths hybridization between the two edges gives rise to a charge transfer with a slight deviation of the free charge density from its asymptotic limit and thus with the appearance of a small electric field.

\begin{figure}
\includegraphics{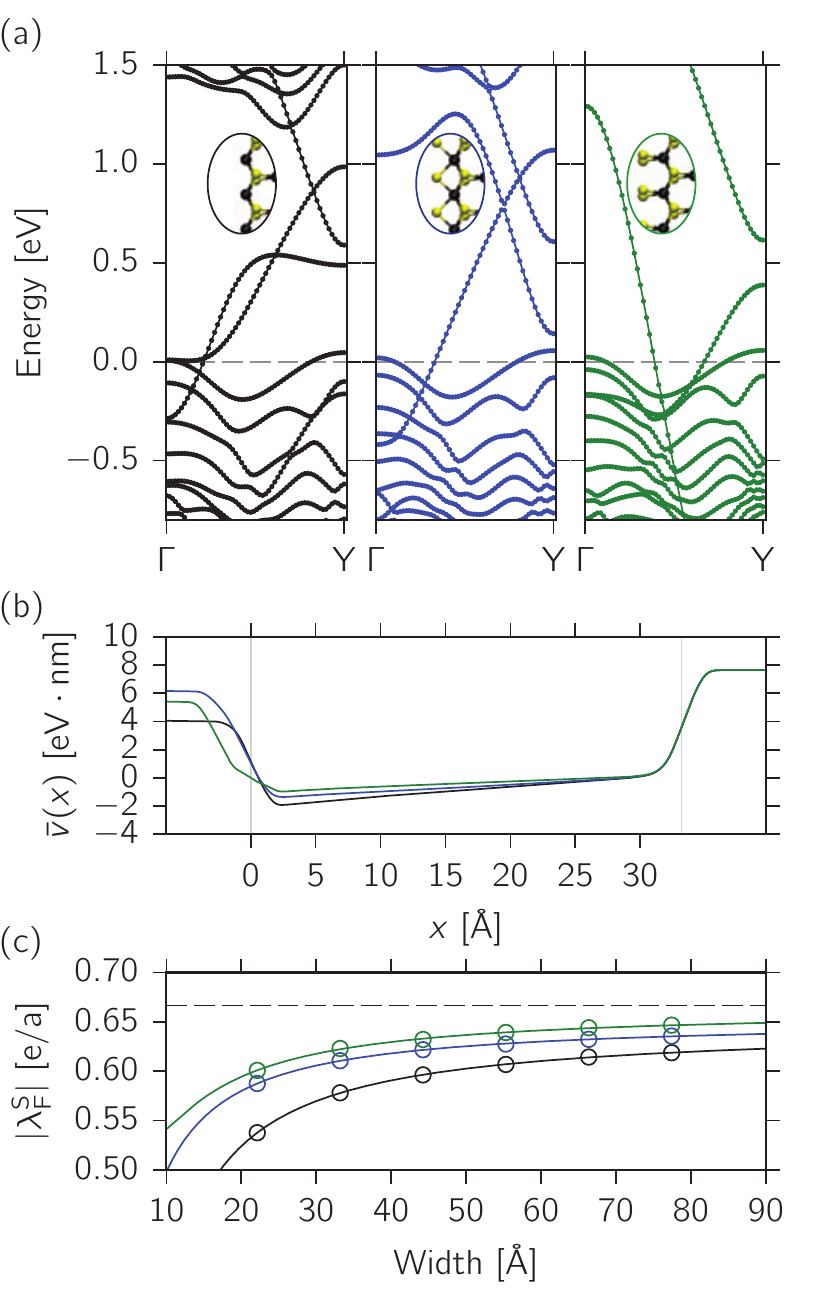}
\caption{Panel (a): Energy band structures for MoS$_2$ nanoribbons ($N_{\rm w}=6$) with different sulfur content on the Mo-edge: bare edge (black), S monomers (blue) and S dimers (green).
Panel (b): Macroscopic average of the electrostatic potential energy defined in Eq.~(\ref{eq:avpot}) as a function of the coordinate across the ribbons introduced in panel (a). 
Panel (c): Charge density of free carriers on the S-edge, $|\lambda^{\rm S}_{\rm F}|$, as a function of the nanoribbon width for different Mo-edge terminations. 
Lines are fits according to Ref.~\onlinecite{Gul2013}.
The color coding in panels (b) and (c) is the same as in panel (a).\label{fig:sulfur}}
\end{figure}

We now turn our attention to the case when additional sulfur atoms are present on the Mo-edge while the S-edge remains pristine. This situation is extremely relevant for experiments since the bare Mo-edge is not stable over a wide range of sulfur chemical potentials\cite{Bol2003,Sch2002}, while the bare S-edge is favored with respect to edge hydrogenation or desulfurization\cite{Bol2003}. In order to restore the bulk six-fold coordination of Mo atoms and stabilize the Mo-edge, it is energetically favorable to bind additional S atoms. Two plausible configurations have been identified\cite{Sch2002,Bys1999}, which differ in the sulfur content at the edge, i.e. in the number of S atoms adsorbed per unit length. One is the S dimer configuration that results from the adsorption of two additional S atoms per unit length that bind to the same Mo atom, similarly to what happens in the bulk. Another is the S monomer configuration that corresponds to the adsorption of a single S atom per unit length, which lies in the Mo-plane and binds to two neighboring external Mo atoms. 
The sulfur content at the edge with respect to the bulk is thus 100\% for S dimers, 50\% for S monomers, and 0\% for bare edges.

For S monomers, since the S atom has a larger electron affinity than Mo, its presence modifies the edge charge reorganization described above for bare edges. In this case the two electrons that would be associated with a Wannier function centered outside the ribbon are taken by the S monomer rather than by the outer Mo atom. This does not change the total amount of bound charge, which is then simply given by the polarization charge, $\lambda^{\rm Mo}_{\rm b}=\lambda_{\rm P}^{\rm Mo}$, as for bare edges. As a consequence, there is still an electric field that drives an insulator-to-metal transition and asymptotically the charge density of free carriers will be the same as for bare edges. This is confirmed by the results of Fig.~\ref{fig:sulfur} where it is shown that: (i) nanoribbons with S-monomer terminated Mo-edges are still metallic; (ii) from the finite slope of $\bar v(x)$ one can infer the presence of a finite electric field; (iii) the asymptotic limit of $|\lambda_{\rm F}|$ is still given by $2e/(3a)$.

In the case of the S-dimer configuration, the bond between S atoms leads to a reduction in the electron affinity of the dimer with respect to atomic sulfur. As a consequence, the dimer is able to only partially transfer the electrons that for bare edges would be captured by the outer Mo atom. This means that the $+2e$ bound charge per unit length associated with the empty Wannier function is no longer fully screened by the negative charge taken by the S dimer. Thus, one has that $\lambda_{\rm b}^{\rm Mo} = \lambda_{\rm P}^{\rm Mo}+2e/a=8e/(3a)$, while at the S-edge $\lambda_{\rm b}^{\rm S}=\lambda_{\rm P}^{\rm S}=-2e/(3a)$. An electric field due to these bound charges is still present,  as shown in Fig.~\ref{fig:sulfur}(b), and points in the same direction as for the case of bare edges. We note that at the Mo-edge the spatial separation between the partially screened polarization charge and the negative charge accumulated on the S dimer gives rise to a finite dipole signaled by a locally negative slope in $\bar v(x)$. A charge reconstruction occurs to screen the bound charges and the system becomes metallic (see panel (a) in Fig.~\ref{fig:sulfur}). In agreement with Ref.~\onlinecite{Bol2003}, three bands cross the Fermi energy: one is associated with the metallic states on the S-edge and it is unaffected by the presence of the dimers; the other two are related to  metallic states on the Mo-edge, one having contributions mainly from the S dimer while the other from outmost Mo atoms.  In this case the asymptotic limit of the free charge density is different at the two edges since the bound charge is different, i.e. $|\lambda_{\rm F}^{\rm Mo}|$ approaches $8e/(3a)$ asymptotically, while $|\lambda_{\rm F}^{\rm S}|\to2e/(3a)$.

\begin{figure}
\includegraphics{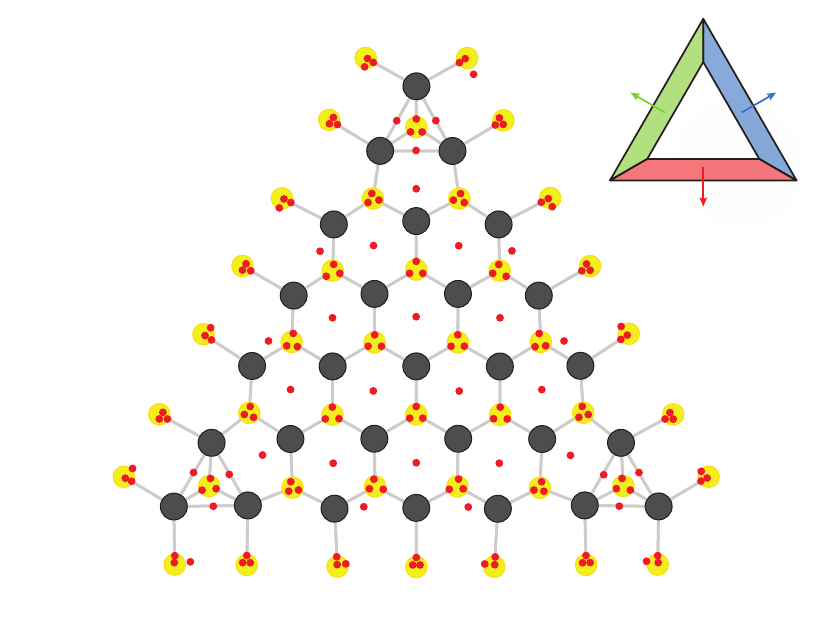}
\caption{Top view of a 112-atom MoS$_2$ triangular island with zigzag Mo-edges saturated by S dimers. As in Fig.~\ref{fig:wannier}, Mo and S atoms are represented as dark grey and light yellow circles, respectively, while small red circles denote the positions of the Wannier functions centers associated with the occupied electronic states. 
Inset: Schematic representation of the triangular island with vectors representing the directions of the three possible polarization vectors with smallest magnitude which are compatible with the bulk polarization lattice in Eq.~(\ref{eq:pol_tmd}). While in the interior (white region) all vectors are equivalent, close to the edges (colored regions) the presence of a specific termination constrains the polarization to a single physical value.
\label{fig:triangles}}
\end{figure}

What has been discussed so far for zigzag nanoribbons is also key to understand the presence of edge states also in triangular islands~\cite{Bol2001,Bol2003}. We show in Fig.~\ref{fig:triangles}  the relaxed structure of a 112-atom MoS$_2$ triangular island with zigzag Mo-edges saturated by S dimers. We will focus on this edge structure since according to previous calculations it is the most stable at standard experimental conditions~\cite{Bol2003,Sch2002}, even though the following arguments apply also to different edge terminations. 
Since we are now dealing with a finite-dimensional system we can define the total electric dipole, which can be computed e.g. by mapping the occupied states into maximally-localized Wannier functions as was done in Eq.~(\ref{eq:pol_wannier}).  The centers of these Wannier functions centers are reported in Fig.~\ref{fig:triangles} as small red circles. In agreement with three-fold rotation symmetry, the total electric dipole vanishes. Nonetheless, this does not mean that locally the electric dipole is identically zero. In addition, we note that in the interior of the triangular island the Wannier functions's centers are in the same positions as in bulk MoS$_2$ (see Fig.~\ref{fig:wannier}). From the earlier discussion, we recall that the bulk polarization of group-VI TMDs is always non-zero, and is given by a lattice of vectors generated by letting ${\bm R}$ range over all direct lattice vectors in Eq.~(\ref{eq:pol_wannier}). In particular, this lattice includes three possible polarization vectors with the smallest magnitude, whose directions with respect to the island edges are represented with vectors of different colors in the inset of Fig.~\ref{fig:triangles}. While in the bulk crystal or in the interior part of the island all these polarization vectors are equivalent, the presence of an edge breaks the translational invariance and thus the equivalence between the different polarizations (see Ref.~\citenum{Mel2002} for a related discussion in the case of boron-nitride nanotubes). Indeed, a specific edge termination reduces the freedom in the choice of the bulk unit cell and thus constrains the possible values of the polarization which are consistent with the crystal termination\cite{King1992,Ste2011}. This was implicitly assumed in Eq.~(\ref{eq:pol_tmd}) where a specific value of the polarization was adopted according to the nanoribbon termination. In the case of triangular islands we have three different edge orientations, each compatible with a different polarization vector which is orthogonal to the edge. We can thus say that, while in the interior of the triangular island all polarization vectors are admissible, close to the edges only a specific polarization emerges as schematically represented in the inset of Fig.~\ref{fig:triangles}.
This preserves the zero total electric dipole but gives rise to a finite (positive) polarization charge at the edge of a triangular island. A charge reconstruction then occurs in order to screen this polarization charge (together with possible additional bound charges) with the accumulation of electrons close to the edge and the emergence of mid-gap states close to the chemical potential. Thus, also in the case of triangular islands, we can explain the existence of such mid-gap edge states (already identified both theoretically\cite{Bol2003} and experimentally\cite{Bol2001}) as a consequence of the polar discontinuity occurring at its boundaries. 
A connection with the case of nanoribbons can be drawn in the asymptotic limit of infinitely large triangular islands, when each edge is independent of the others and the system becomes metallic with a continuous set of mid-gap states localized close to the edges and with a charge equal and opposite to the bound charge.

\begin{figure*}
\includegraphics{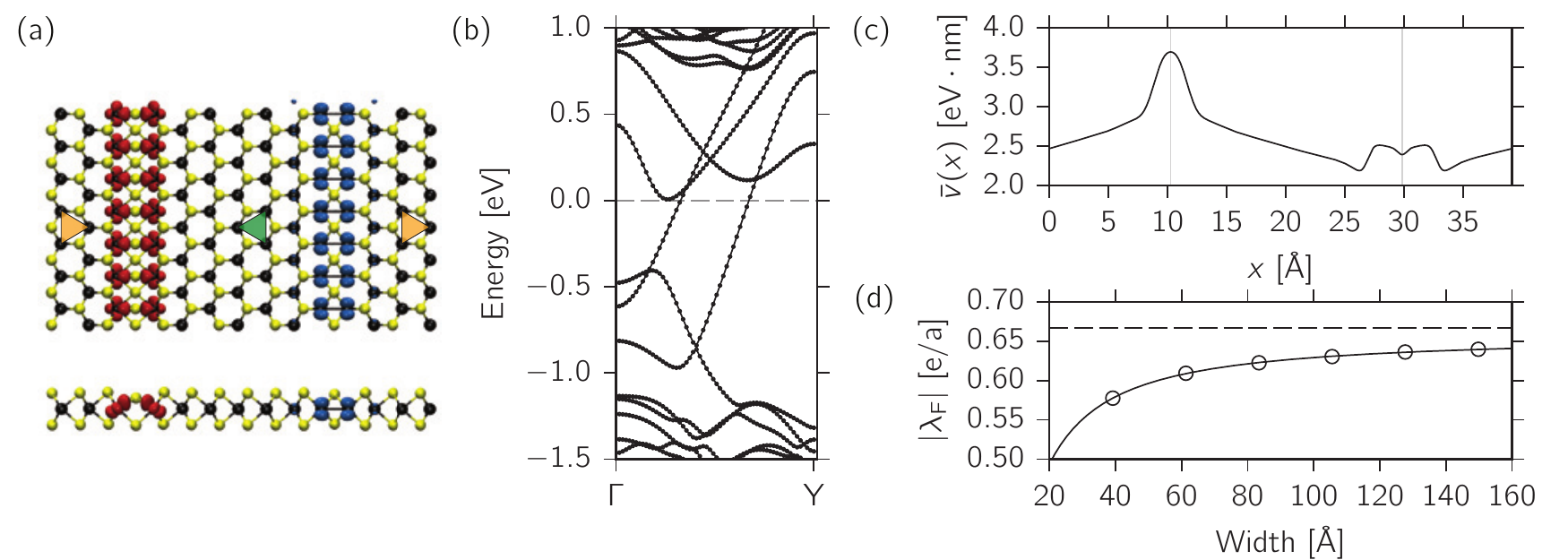}
\caption{
Panel (a): Top and lateral views of a MoS$_2$ structure containing two inversion domain boundaries extending along a zigzag direction. Green and orange triangles  emphasize the opposite atomic arrangement at each side of the inversion domain boundaries. The real-space distribution of the charge density of free electrons (blue) and holes (red) is superimposed. 
Panel (b): Energy band structure as a function of the crystal momentum along the interface for the inversion domain boundary shown in panel (a).The dashed line highlights the position of the Fermi energy.
Panel (c): Macroscopic average of the electrostatic potential energy defined in Eq.~(\ref{eq:avpot}) as a function of the coordinate across the system. Thin solid lines identify the position of the two inversion domain boundaries in the supercell.  
Panel (d): Charge density of free carriers $|\lambda_{\rm F}|$ as a function of the total width. The dashed line  highlights the asymptotic limit $2e/(3a)$.
The solid line is a fit according to Ref.~\onlinecite{Gul2013}.
\label{fig:twin}}
\end{figure*}

Zigzag nanoribbons and triangular islands are not the only expression of metallicity in group-VI TMDs. Indeed, as we mentioned above, there is both theoretical\cite{Zou2013,Zhou2013} and experimental\cite{Liu2014} evidence that metallic states appear also at zigzag inversion domain boundaries (IDBs). These grain boundaries emerge as a result of the lack of inversion symmetry in group-VI TMD monolayers when two crystallites related by inversion are merged together. 
An explanation of the origin of metallicity in zigzag IDB is still missing. In the following we show that when the IDB extends along a zigzag direction a polar discontinuity emerges across the boundary where polarization charges appear. Boundary states are then induced to screen these polarization charges and the system becomes metallic. 
In order to verify this picture within first-principles DFT simulations we considered supercells containing two IDBs  extending along a zigzag direction and separating two stripes of MoS$_2$ with opposite atomic arrangement. We assumed the structure of the S-boundary to be the one identified in experiments\cite{Zhou2013} (see the right boundary in Fig.~\ref{fig:twin}(a)). 
In order to assess the structure of the Mo-boundary we performed several calculations for different atomic arrangements, assuming both stoichiometric and non-stoichiometric configurations. In agreement with Ref.~\onlinecite{Zou2013}, we obtained that the stoichiometric Mo-boundary structure shown in Fig.~\ref{fig:twin}(a) is the most stable over a wide range of sulfur chemical potentials and we will assume it in the following. We note that for this structure both the S- and Mo-boundaries display reflection symmetry through a vertical plane containing the interface. 
The energy bands as a function of momentum along the IDBs are plotted in Fig.~\ref{fig:twin}(b) and are consistent with the results reported in Ref.~\onlinecite{Zou2013}. In particular, we note that the system is metallic with two bands crossing the Fermi energy: one is a valence band that gets partially depleted while the other is a partially-filled conduction band. The charge density of free carriers associated with these two bands is shown in Fig.~\ref{fig:twin}(a), with holes localized on one IDB and electrons on the opposite IDB. Metallicity is induced by the electrostatic potential --whose macroscopic average $\bar v(x)$  is shown in Fig.~\ref{fig:twin}(c)-- that shifts the bands upwards in energy at one IDB creating a hole pocket, while downwards at the opposite IDB giving rise to an electron pocket. 
To understand the origin of this electrostatic potential we first recall that because of the opposite orientation of the crystals facing the IDB, we have a polar discontinuity across the boundary. For an IDB extending along a zigzag direction, we have that the density of polarization charges appearing at the boundary is given by $\lambda_{\rm P}=\pm4e/(3a)$, where the positive (negative) sign holds for the Mo(S)-boundary. The signs of these charges are not consistent with the change in slope of $\bar v(x)$ in Fig.~\ref{fig:twin}(c), suggesting that there must be additional contributions to the bound charge at the boundaries. We first note that at the S-boundary, the vertical plane of S atoms is shared between the crystals at the two sides so that the calculation of the polarization charges based on the bulk properties leads to a double counting of the contribution from the S atoms ($2\times2e/a$). In addition, the formation of a Mo-Mo bond leads to the localization of two additional electrons at the boundary ($-2e/a$). This means that the bound charge can not be identified with the polarization charge reported above but has to be corrected in the following way: $\lambda_{\rm b}^{\rm S} = \lambda_{\rm P}^{\rm S} +4e/a -2e/a = +2e/(3a)$. 
At the Mo-boundary, we have  instead an extra row of S atoms which are bonded to the terminal Mo atoms at both sides. Owing to the different S-coordination with respect to the bulk, six electrons are needed to form such bonds. Four of them come from a charge reorganization at the interface and would otherwise be associated with Wannier functions localized at the boundary. The last two supplementary electrons give rise to an additional $-2e/a$ contribution to the bound charge: $\lambda_{\rm b}^{\rm Mo} = \lambda_{\rm P}^{\rm Mo}-2e/a=-2e/(3a)$. 
We thus have that the bound charge has the same magnitude at both boundaries and changes sign with respect to the bare polarization contribution. This is now consistent with the change in slope of $\bar v(x)$ and explains the origin of the electrostatic potential. 
The free carriers appearing at the boundaries have to asymptotically screen the bound charge in order to prevent a divergency in the electrostatic energy. In Fig.~\ref{fig:twin}(d) we plot the free charge $|\lambda_{\rm F}|$ as a function of the total width of the system (approximately twice the separation between successive IDB). Indeed, the asymptotic limit is consistent with the value of the bound charge obtained above, $|\lambda_{\rm F}|\to2e/(3a)$. We have thus identified a consistent picture to rationalize the presence of metallic states previously reported at IDBs\cite{Zou2013,Zhou2013,Liu2014}.  

Before concluding we stress that these 1D wires of free carries in group-VI TMD nanostructures have the potential to deliver extremely promising applications. In addition to their established role in catalysis and their relevance as 1D channels in ultra-thin electronics and spintronics, we envision fruitful applications in solar-energy harvesting. Indeed, these systems support an intrinsic photovoltaic effect: an electron-hole pair created in the ``bulk'' semiconducting interior of such TMD nanostructures is split by the built-in electric field due to the polarization charges. The electron and the hole are then driven towards opposite interfaces where they can be collected at the 1D wires of free carries, as already discussed in Ref.~\onlinecite{Gib2014}.

In conclusion, we have shown with extensive first-principles DFT simulations that metallicity in TMD nanostructures is associated with a widespread and universal phenomenon driven by polar discontinuities emerging across zigzag edges or inversion domain boundaries and it is robust over different specific edge/boundary configurations. 
The charge density of free carriers increases with the size of the system until asymptotically it perfectly screens the bound charges at the edges/boundaries in order to prevent the divergence of the electrostatic potential. 
The polarization charges associated with the polar discontinuity give the major contribution to the bound charge at each edge/boundary. Nonetheless, the knowledge of the bulk polarization alone is not sufficient to predict the asymptotic value of the free charge density as additional bound charges can appear depending on the edge/boundary structure and induce more free carriers. This happens for instance when hydrogen atoms or sulfur dimers are chemisorbed to terminate the dangling bonds at nanoribbon edges. 
Even for finite-dimensional nanostructures like triangular islands, where the total electric dipole is zero, a polar discontinuity still occurs across their edges and explains the emergence of mid-gap edge states previously discussed in the literature.

\section*{Methods}
We performed first-principles density-functional-theory simulations using the PWscf code of the Quantum-ESPRESSO distribution\cite{Gia2009}.
The exchange and correlation functional is that of Perdew-Burke-Ernzerhof~\cite{PBE} form of generalized gradient corrections. We use ultrasoft pseudopotentials\cite{Vand1990} from {\tt PSlibrary.0.2.5}\cite{pslibrary} with energy cutoffs of 60 and 600 Ry for wave functions and density, respectively.
Bulk relaxed structures are obtained within the Broyden-Fletcher-Goldfarb-Shanno algorithm by requiring that the forces acting on atoms are below 1~meV$/$\AA\ and the residual stress on the cell is less than 0.5 kbar. Nanoribbons are then obtained by breaking the periodicity along a given direction without further relaxation; only the position of H or S atoms saturating the edge bonds are relaxed in this case. We checked that including the atomic relaxations for all other atoms in the nanoribbons leads only to marginal changes in the charge density of free carriers; only for extremely narrow ones ($N_{\rm w}<3$ for MoS$_2$) structural reconstructions can occur as a competing mechanism to compensate polarity\cite{Gul2015}. Finally, for inversion domain boundaries we relaxed the positions of atoms in the first rectangular unit cells close to the boundary while keeping the bulk periodicity along the interface.
For nanoribbons and inversion domain boundaries a fine $1\times18\times1$ Monkhorst-Pack grid coupled to a small cold smearing\cite{Mar1999} (0.001 Ry) are used to sample the Brillouin zone and to obtain accurate results for the free carrier charge density.
In order to get rid of the spurious interactions with periodic replicas along the vertical direction, we correct the electrostatic potential using the Otani-Sugino approach\cite{Ota2006} with 20~\AA\ of vacuum. For nanoribbons,  spurious periodic replicas are present also in-plane. In this case their electrostatic interaction is reduced by allowing for an amount of vacuum equal to the nanoribbon width ($>20$~\AA).
The charge density of free carriers $\lambda_{\rm F}$ is  the amount of charge per unit length by which valence/conduction bands are depleted/filled and is computed by integrating the corresponding density of states.

\section*{Acknowledgements}
We would like to acknowledge many useful discussions with Giovanni Pizzi, Francesco
Mauri, and Stefano Baroni. We also acknowledge partial support from the Graphene Flagship
initiative and simulation time from the Swiss National Supercomputing Centre (CSCS)
through project ID s337.

\end{document}